\documentclass[12pt]{article}
\usepackage{amsfonts}
\usepackage{amssymb}
\usepackage{amsfonts}
\usepackage{amsmath}
\usepackage{epsfig}
\usepackage{color}

\newcommand{\mc}{\mathcal}
\newcommand{\Hil}{\mc H}
\newcommand{\D}{{\mc D}}

\newcommand{\be}{\begin{equation}}
\newcommand{\en}{\end{equation}}
\newcommand{\bea}{\begin{eqnarray}}
\newcommand{\ena}{\end{eqnarray}}
\newcommand{\beano}{\begin{eqnarray*}}
\newcommand{\enano}{\end{eqnarray*}}
\newcommand{\bee}{\begin{enumerate}}
\newcommand{\ene}{\end{enumerate}}

\newcommand{\F}{{\cal F}}

\begin{document}

\thispagestyle{empty}

\vspace*{2cm}

\begin{center}
{{\Large \bf Generalized Bogoliubov transformations versus ${\mathcal{D}}$%
-pseudo-bosons}}\\[10mm]


{\large F. Bagarello} \footnote[1]{ Dipartimento di Energia, Ingegneria dell'Informazione e Modelli Matematici,
Facolt\`a di Ingegneria, Universit\`a di Palermo, I-90128  Palermo, and INFN, Universit\`a di di Torino, ITALY\\
e-mail: fabio.bagarello@unipa.it\,\,\,\, Home page: www.unipa.it/fabio.bagarello}

{\large A. Fring} \footnote[2]{Department of Mathematics, City University London,\\
$\,\,$ Northampton Square, London EC1V 0HB, UK\\
e-mail: a.fring@city.ac.uk}

\end{center}

\vspace*{2cm}

\begin{abstract}
\noindent We demonstrate that not all generalized Bogoliubov transformations lead to $\D$-pseudo-bosons and prove that a correspondence between the two can only be achieved with the imposition of specific constraints on the parameters defining the transformation. For certain values of the parameters we find that the norms of the vectors in sets of eigenvectors of two related apparently non self-adjoint number-like operators possess different types of asymptotic behavior. We use this result to deduce further that they constitute bases for a Hilbert space, albeit neither of them can form a Riesz base. When the constraints are relaxed they cease to be Hilbert space bases, but remain $\D$-quasi bases.

\end{abstract}


\vfill

\section{Introduction}

In this manuscript we compare two concepts which have facilitated the study
on non-Hermitian quantum systems in recent years, \emph{generalized
Bogoliubov transformations} (GBTs) and $\mathcal{D}$-\emph{pseudo-bosons} ($%
\mathcal{D}$-PBs). Both ideas can be employed to address different aspects
of the key questions in the study of non-Hermitian quantum systems: Under
which circumstances are the spectra of non-Hermitian Hamiltonians real and
what kind of metric needs to be employed to render the system physically
meaningful? These issues have been the subject of investigations since the
seminal papers \cite{sgh,ben} on this topic and aspects of them have been
answered for many different types of models. The underlying principle of
both ideas make use of simple deformations of the canonical commutation
relation $[c,c^{\dagger }]=1\!\!1$. This principle restricts, of course, our analysis to a subset of the class of $PT$-quantum mechanics which have been studied, in recent years, by several authors, see \cite{benrev,miloshinbagbook} and references therein. However, this subset is rather large and contains several well known systems, \cite{baginbagbook}, like the Swanson model, just to cite one. 

Bogoliubov transformations are linear transformations mapping the operators $%
c$ and $c^{\dagger }$ to a new canonical pair $a$ and $b$. They were
originally introduced to aid the study of pairing interaction in
superconductivity \cite{Bogo} and have been generalized thereafter in
various ways, e.g. \cite{GenBog}. When the operators in the new pair are not
assumed to be mutually adjoint, i.e. $b^{\dagger }\neq a$, these maps are
usually referred to as generalized Bogoliubov transformations. In the
context of the study of non-Hermitian quantum systems they were employed to
establish the reality of the spectra of certain non-Hermitian Hamiltonians
and to identify well-defined metric operators which map the system to
isospectral Hermitian counterparts \cite{Swanson,andreas1}.

Domain issues are often left unaddressed in these constructions.
They also constitute interesting mathematical problems in their own right
and spectral properties of non self-adjoint operators can be quite
intricate, see for example \cite{goh,davies} or the more recent volume \cite%
{baginbagbook}. One of the mathematical difficulties of non self-adjoint Hamiltonians
is related to the eigenvectors $\varphi _{n}$ of $H$ and $\Psi _{n}$ of $%
H^{\dagger }$ for $n\geq 0$, if they exist. One can not simply
assume that the sets of these eigenvectors constitute biorthogonal bases of
the Hilbert space $\mathcal{H}$ on which the models are defined. One needs
to verify this property in detail and indeed in various models this
assumption has turned out to be incorrect, as discussed in \cite{davies2}-\cite{bagfring}, for instance. In
order to understand these aspects in depth a large series of investigations
has been carried out in recent years \cite{bag1,bag2,bag3,fb1,bagfring} on
so-called \emph{pseudo-bosonic} systems \cite{tri}. The obtained results
were recently reviewed and extended in \cite{baginbagbook}. In all the explicit
examples studied so far, the eigenvectors of $H=ba$ and $H^{\dagger
}=a^{\dagger }b^{\dagger }$ are biorthogonal, of course, but they are not
bases for $\mathcal{H}$. However, see below, they still produce a weaker
version of the closure relation on some dense subspace $\mathcal{G}$ of $%
\mathcal{H}$ and are therefore coined $\mathcal{G}$-quasi bases.

At first sight GBTs and PBs appear to be quite similar and the natural
question arises under which conditions they might be the same or more
specifically: When do GBTs correspond to PBs? We will demonstrate here that
the latter only happens under suitable conditions. Thus, in general these
two notions are not equivalent. Interestingly, GBTs allow us to find
examples of number-like pseudo-bosonic operators whose eigenstates form
biorthogonal bases when certain requirements are fulfilled and examples in
which this is not true, even if they still provide useful weaker versions of
the resolution of the identity. A specific version of the Swanson model
discussed in \cite{bag3} is an example based on GBT which admits no bases.

Our manuscript is organized as follows: In section 2 we recall some
definitions and general features on GBTs and PBs relevant for our
investigations. In section 3 we analyze in detail under which conditions GBT
give rise to PBs. By imposing suitable constraints on the parameters
defining the GBT, we show that the eigenstates of the operators $N=ba$ and $%
N^{\dagger }$ are indeed biorthogonal, with simple eigenvalues, but they are bases only if these
specific constraints are satisfied. The details are contained in section \ref%
{sect23}. In section \ref{sect24} we show that, if these constraints are not
satisfied, these eigenstates are $\mathcal{G}$-quasi bases but not bases. We
state our conclusions and a further outlook into open problems in section 4.

\section{Generalities and definitions}

\label{sect_fr}

To establish our conventions let us first recall some well-known facts and
definitions about GBTs and PBs.

\subsection{Generalized Bogoliubov transformations\label{secGBT}}

We consider two operators $c$ and $c^{\dagger }$ satisfying the canonical
commutation relation $[c,c^{\dagger }]=1\!\!1$. Taking for instance $c=\frac{%
1}{\sqrt{2}}\left( x+\frac{d}{dx}\right) $ and $c^{\dagger }=\frac{1}{\sqrt{2%
}}\left( x-\frac{d}{dx}\right) $ is a well-known possible realization of
these operators in the Hilbert space $\mathcal{H}=\mathcal{L}^{2}(\mathbb{R)}
$ of the square integrable functions on $\mathbb{R}$. We use here the
convention $\hbar =1$ when comparing to a quantum mechanical setting.

GBTs \cite{Swanson,andreas1} are linear maps defined as
\begin{equation}
\left(
\begin{array}{c}
a \\
b%
\end{array}%
\right) =\left(
\begin{array}{cc}
\beta  & -\delta  \\
-\alpha  & \gamma
\end{array}%
\right) \left(
\begin{array}{c}
c \\
c^{\dagger }%
\end{array}%
\right) ,  \label{21}
\end{equation}%
or more explicitly, the new operators are $a=\beta c-\delta c^{\dagger }$
and $b=-\alpha c+\gamma c^{\dagger }$. Here $\alpha $, $\beta $, $\gamma $, $%
\delta \in \mathbb{C}$ and to ensure that $[a,b]=1\!\!1$ we require $\det
(T)=\beta \gamma -\alpha \delta =1$, where $T$ is the two-by-two matrix in
the defining relation for the transformation (\ref{21}). In addition, we
restrict here to the choices of $\alpha $, $\beta $, $\gamma $ and $\delta $
such that $b\neq a^{\dagger }$. Since $\det (T)=1\neq 0$, the inverse GBT
always exists
\begin{equation}
T^{-1}=\left(
\begin{array}{cc}
\gamma  & \delta  \\
\alpha  & \beta
\end{array}%
\right) .
\end{equation}

PBs are very similar objects as the operators $a$ and $b$ produced by the
GBT. In fact, they were formally (i.e., with no care on the domains of the unbounded operators involved in their framework, as well as on other mathematical subtleties) introduced in \cite{tri}, and then in a more rigorous way in \cite{bag1}, by taking the
commutation relation $[a,b]=1\!\!1$ between two different operators $a$ and $%
b$ densely defined on a Hilbert space $\mathcal{H}$ as the primary object
and it was found that when $b\neq a^{\dagger }$ interesting situations
arose. When one wishes to apply these operators in building non-Hermitian
quantum mechanical models further constraints are needed. To allow the usage
of $a$ and $b$ for the construction of two families of biorthogonal vectors
of $\mathcal{H}$ some specific scenarios were dealt with in \cite{bag1}.
Also when considering non-Hermitian models constructed from $\mathcal{PT}$%
-invariant combinations of $c$ and $c^{\dagger }$, the requirement that they
can be mapped by means of GBT into a form of a harmonic oscillator plus a
Casimir operator \cite{andreas1} imposes further constraints on the
constants $\alpha $, $\beta $, $\gamma $, $\delta $.

Next we recall some definitions and relevant facts for reference about PBs.

\subsection{Pseudo-bosons}

\label{subDPBs}

\textbf{Definition }${\mathcal{D}}$\emph{-}\textbf{PB:} \label{FBdef21}
\emph{The pair of operators }$a$\emph{\ and }$b$\emph{\ are called }${%
\mathcal{D}}$\emph{-pseudo-bosons (}${\mathcal{D}}$\emph{-PB) if, for all }$%
f\in {\mathcal{D}}$\emph{, we have }%
\begin{equation}
a\,b\,f-b\,a\,f=f.  \label{FB31}
\end{equation}

\emph{The domain }$\mathcal{D}$\emph{\ is a dense subspace of a Hilbert
space }$\mathcal{H}$\emph{ stable under the action of }$a$\emph{, }$b$\emph{%
, }$a^{\dagger }$\emph{\ and }$b^{\dagger }$\emph{, that is  }$a^{\sharp }\,{\mathcal{%
D}}\subseteq {\mathcal{D}}$\emph{\ and }$b^{\sharp }\,{\mathcal{D}}\subseteq {%
\mathcal{D}}$\emph{, where }$x^{\sharp }$\emph{\ is }$x$\emph{\ or }$%
x^{\dagger }$\emph{. }

Note that since $a^{\sharp }f$ is well defined and belongs to ${\mathcal{D}}$
for all $f\in {\mathcal{D}}$, it is clear that ${\mathcal{D}}\subseteq
D(a^{\sharp })$, the domain of the operator $a^\sharp$. The analogue holds also for ${\mathcal{D}}\subseteq
D(b^{\sharp })$. We often use a simplified notation and instead of (\ref%
{FB31}) we only write $[a,b]=1\!\!1$, where, as before, $1\!\!1$ is the identity
operator on $\mathcal{H}$, keeping in mind that both sides of this equation
have to act on a certain $f\in \mathcal{D}$.

In addition we assume:

\noindent \textbf{Assumption ${\mathcal{D}}$-PB 1:} There exists a non-zero $%
\varphi _{0}\in {\mathcal{D}}$ such that $a\,\varphi _{0}=0$.

\noindent \textbf{Assumption ${\mathcal{D}}$-PB 2:} There exists a non-zero $%
\Psi _{0}\in {\mathcal{D}}$ such that $b^{\dagger }\,\Psi _{0}=0$.

\noindent \textbf{Assumption ${\mathcal{D}}$-PB 3: }The set $\mathcal{F}%
_{\varphi }:=\{\varphi _{n},\,n\geq 0\}$ is a basis for $\mathcal{H}$.

Note that the vectors
\begin{equation}
\varphi _{n}:=\frac{1}{\sqrt{n!}}\,b^{n}\varphi _{0},\qquad \Psi _{n}:=\frac{%
1}{\sqrt{n!}}\,{a^{\dagger }}^{n}\Psi _{0},  \label{FB32}
\end{equation}%
are well-defined for $n\geq 0$, since ${\mathcal{D}}$ is stable under the
action of $b$ and $a^{\dagger }$. In particular, it is obvious that $\varphi
_{0}\in D^{\infty }(b):=\cap _{k\geq 0}D(b^{k})$ and that $\Psi _{0}\in
D^{\infty }(a^{\dagger })$. Therefore we can introduce the sets $\mathcal{F}%
_{\varphi }$ and in addition $\mathcal{F}_{\Psi }=\{\Psi _{n},\,n\geq 0\}$.
By the same reasoning we also deduce that each $\varphi _{n}$ and each $\Psi
_{n}$ belongs to ${\mathcal{D}}$ and therefore to the domains of $a^{\sharp
} $, $b^{\sharp }$ and $N^{\sharp }$, where $N:=ba$, $N^{\dagger
}=a^{\dagger }b^{\dagger }$.

The following lowering and raising relations are then easily obtained%
\begin{equation}
\begin{array}{llll}
~a\varphi _{n}=\sqrt{n}\,\varphi _{n-1},~a\,\varphi _{0}=0,~~~ &  &
b^{\dagger }\Psi _{n}=\sqrt{n}\,\Psi _{n-1},~b^{\dagger }\Psi _{0}=0,~~ &
\text{for }n\geq 1,\text{ } \\
a^{\dagger }\Psi _{n}=\sqrt{n+1}\,\Psi _{n+1}, &  & ~b\,\varphi _{n}=\sqrt{n+1}\,%
\varphi _{n+1}, & \text{for }n\geq 0,\text{ }%
\end{array}
\label{33}
\end{equation}%
as well as the eigenvalue equations $N\varphi _{n}=n\varphi _{n}$ and $%
N^{\dagger }\Psi _{n}=n\Psi _{n}$ for $n\geq 0$. Notice that all the eigenvalues are simple.  As a consequence of these last
equations we derive that
\begin{equation}
\left\langle \varphi _{n},\Psi _{m}\right\rangle =\delta _{n,m},
\label{FB34}
\end{equation}%
for all $n,m\geq 0$, when choosing the normalization of $\varphi _{0}$ and $%
\Psi _{0}$ in such a way that $\left\langle \varphi _{0},\Psi
_{0}\right\rangle =1$. Then $\mathcal{F}_{\varphi }$ and $\mathcal{F}_{\Psi }
$ are biorthonormal sets of eigenstates of $N$ and $N^{\dagger }$,
respectively.

The assumptions $\mathcal{D}$-PB 1 and $\mathcal{D}$-PB 2 do in principle
not allow to conclude anything about the fact that $\mathcal{F}_{\varphi }$
and $\mathcal{F}_{\Psi }$ are also bases for $\mathcal{H}$, or even whether
they are Riesz bases, which is the reason for making the assumption $%
\mathcal{D}$-PB 3. Notice that it automatically implies that $\mathcal{F}%
_{\Psi }$ is a basis for $\mathcal{H}$ as well \cite{you}. However, during
the years several examples in which this natural assumption is not satisfied
have been found, see for instance \cite{baginbagbook} and references therein. For
this reason a weaker version of assumption ${\mathcal{D}}$-PB 3 was
introduced in \cite{bag2}.

\noindent \textbf{Assumption ${\mathcal{D}}$-PBw 3: }$\mathcal{F}_{\varphi }$
and $\mathcal{F}_{\Psi }$ are $\mathcal{G}$\emph{-quasi bases} for some
subspace $\mathcal{G}$ dense in $\mathcal{H}$.

Two biorthogonal sets $\mathcal{F}_{\eta }=\{\eta _{n}\in \mathcal{G}%
,\,g\geq 0\}$ and $\mathcal{F}_{\Phi }=\{\Phi _{n}\in \mathcal{G},\,g\geq
0\} $ have been called \emph{$\mathcal{G}$}-quasi bases when
\begin{equation}
\left\langle f,g\right\rangle =\sum_{n\geq 0}\left\langle f,\eta
_{n}\right\rangle \left\langle \Phi _{n},g\right\rangle =\sum_{n\geq
0}\left\langle f,\Phi _{n}\right\rangle \left\langle \eta
_{n},g\right\rangle ,  \label{FB35}
\end{equation}
holds for all $f,g\in \mathcal{G}$. It is clear that assumption ${\mathcal{D}%
}$-PB 3 implies its weaker version (\ref{FB35}), but the reverse can not be
inferred. However, when $\mathcal{F}_{\eta }$ and $\mathcal{F}_{\Phi }$
satisfy the relation (\ref{FB35}) we still have some (weak) form of
resolution of the identity and from a physical point of view we are still
 able to deduce interesting results, \cite{bag2}. Incidentally we see that if $f\in
\mathcal{G}$ is orthogonal to all the $\Phi _{n}$'s (or to all the $\eta
_{n} $'s), then $f$ is necessarily zero. Hence, both $\mathcal{F}_{\eta }$
and $\mathcal{F}_{\Phi }$ are automatically complete in $\mathcal{G}$. For
further results on $\mathcal{G}$-quasi bases we refer the reader to \cite%
{bag2}, where a discussion can be found in which sense these bases extend
Riesz biorthogonal bases and additional results on the mathematical
structure arising out of $a$, $b$ and the various vectors introduced so far.

Here we will be mainly interested in demonstrating the interesting fact that
depending on the choices of parameters involved in the GTBs they provide
examples in which $\mathcal{F}_{\varphi }$ and $\mathcal{F}_{\Psi }$ are
indeed bases such that assumption ${\mathcal{D}}$-PB 3 holds as well as
examples in which they are just $\mathcal{G}$-quasi bases for some dense $%
\mathcal{G}\subset \mathcal{H}$, so that assumption ${\mathcal{D}}$-PBw 3
holds while assumption ${\mathcal{D}}$-PB 3 does not.

\section{GBTs versus PBs}

We are now in the position to address the following questions: \emph{does a
GBT always produce ${\mathcal{D}}$-PBs?} or more specifically \emph{when
does a GBT produce ${\mathcal{D}}$-PBs?} In fact, the first question can
already be answered negatively by several simple counter examples. For instance,  already in \cite{tri,bag1}, it
was shown that specific operators of the form (\ref{21}) satisfying $%
[a,b]=1\!\!1$ need not be pseudo-bosonic. Let us now make this more
evident by treating the general case. We adopt the explicit realization of $%
c $ and $c^{\dagger }$ quoted at the beginning of subsection \ref{secGBT}.
Hence we obtain%
\begin{eqnarray}
a &=&\frac{1}{\sqrt{2}}\left[ (\beta -\delta )x+(\beta +\delta )\frac{d}{dx}%
\right] ,\qquad b=\frac{1}{\sqrt{2}}\left[ (\gamma -\alpha )x-(\gamma
+\alpha )\frac{d}{dx}\right] ,  \label{22} \\
a^{\dagger } &=&\frac{1}{\sqrt{2}}\left[ (\overline{\beta }-\overline{\delta
})x-(\overline{\beta }+\overline{\delta })\frac{d}{dx}\right] ,\qquad
b^{\dagger }=\frac{1}{\sqrt{2}}\left[ (\overline{\gamma }-\overline{\alpha }%
)x+(\overline{\gamma }+\overline{\alpha })\frac{d}{dx}\right] .  \label{23}
\end{eqnarray}%
We can easily verify whether the assumptions ${\mathcal{D}}$-PB 1 and ${%
\mathcal{D}}$-PB 2 are satisfied. In particular, the equations for the
ground states $a\varphi _{0}(x)=0$ and $b^{\dagger }\Psi _{0}(x)=0$ admit
the solutions
\begin{equation}
\varphi _{0}(x)=N_{\varphi }e^{-\frac{1}{2}\,x^{2}\,\frac{\beta -\delta }{%
\beta +\delta }},\qquad \Psi _{0}(x)=N_{\Psi }e^{-\frac{1}{2}\,x^{2}\,\frac{%
\overline{\gamma }-\overline{\alpha }}{\overline{\gamma }+\overline{\alpha }}%
},  \label{24}
\end{equation}%
where $N_{\varphi }$ and $N_{\Psi }$ are suitable normalization constants to
be specified further below. The first crucial point to note is that these
two functions do not always belong to $\mathcal{H}=\mathcal{L}^{2}(\mathbb{R)%
}$. This is only true when the following constraints on the parameters in $T$
are satisfied:
\begin{equation}
\Re\left( \frac{\beta -\delta }{\beta +\delta }\right) >0,\qquad \Re\left( \frac{\gamma -\alpha }{\gamma +\alpha }\right) >0.  \label{25}
\end{equation}

Is is evident that (\ref{25}) is distinct from the necessary condition $\det
(T)$ $=1$. It is easily seen that assuming the latter we can produce all
possible scenarios a) $\varphi _{0}(x)\in \mathcal{H}$, $\Psi _{0}(x)\notin
\mathcal{H}$, b) $\varphi _{0}(x)\notin \mathcal{H}$, $\Psi _{0}(x)\in
\mathcal{H}$, c) $\varphi _{0}(x)$, $\Psi _{0}(x)\notin \mathcal{H}$ and d) $%
\varphi _{0}(x)$, $\Psi _{0}(x)\in \mathcal{H}$. Explicit examples for
parameter choices for these cases are for instance a) $\alpha =\gamma
=\delta =1$, $\beta =2$, b) $\alpha =\beta =\delta =1$, $\gamma =2$, c) $%
\alpha =-3/2$, $\beta =1/4$, $\gamma =1$ , $\delta =1/2$ and d) $\alpha =2/3$%
, $\beta =2$, $\gamma =1$, $\delta =3/2$. Thus on the basis of assumptions ${%
\mathcal{D}}$-PB 1 and  ${\mathcal{D}}$-PB 2 alone we conclude
already that the GBT described by the cases a), b) and c) can not be ${%
\mathcal{D}}$-PB. However, the case d) demonstrates that we have GBTs that
might also produce ${\mathcal{D}}$-PBs. Thus we need to verify whether the
remaining assumption ${\mathcal{D}}$-PB 3 in section \ref{subDPBs} also
holds for those cases and of course we also need to fix ${\mathcal{D}}$.

Thus the next step is to compute
\begin{equation}
\varphi _{n}(x)=\frac{1}{\sqrt{n!}}\,b^{n}\varphi _{0}(x),\qquad \Psi
_{n}(x)=\frac{1}{\sqrt{n!}}\,{a^{\dagger }}^{n}\Psi _{0}(x),  \label{def}
\end{equation}%
for $n\geq 0$. It suffices to determine the expression for $\varphi _{n}(x)$
as those for $\Psi _{n}(x)$ can be obtained from the former simply by
replacing $\delta $ with $\overline{\alpha }$ and $\beta $ with $\overline{%
\gamma }$ when noting that $b$ and $\Psi _{0}(x)/N_{\Psi }$ algebraically
coincide with $a^{\dagger }$ and $\varphi _{0}(x)/N_{\varphi }$,
respectively after this requirements. Moreover, with the use of condition (\ref{25}) we can fix the
value for the product of $\overline{N}_{\varphi }$ and $N_{\Psi }$
\begin{equation}
\left\langle \Psi _{0},\varphi _{0}\right\rangle =1\quad \Rightarrow \quad
\overline{N}_{\varphi }N_{\Psi }=\frac{1}{\sqrt{\pi (\alpha +\gamma )(\beta
+\delta )}}.
\end{equation}%
From the definition (\ref{def}) and (\ref{24}) it is easy to see that
\begin{eqnarray}
\varphi _{n}(x) &=&\frac{N_{\varphi }}{\sqrt{n!2^{n}}}\left[ \left( \gamma
-\alpha \right) \,x-(\gamma +\alpha )\frac{d}{dx}\right] ^{n}e^{-\frac{1}{2}%
\,x^{2}\,\frac{\beta -\delta }{\beta +\delta }},  \label{26} \\
&=&\frac{N_{\varphi }}{\sqrt{n!2^{n}}}\left( \frac{\alpha +\gamma }{\beta
+\delta }\right) ^{n/2}H_{n}\left[ \frac{x}{\sqrt{(\alpha +\gamma )(\beta
+\delta )}}\right] \,e^{-\frac{1}{2}\,x^{2}\,\frac{\beta -\delta }{\beta
+\delta }},  \label{25bis}
\end{eqnarray}%
for all $n\geq 0$, with $H_{n}(x)$ denoting the $n$-th Hermite polynomial.
The constraint needed for these functions to be square integrable is the
same as (\ref{25}), since they are simply polynomials times the same
Gaussian that already appeared in $\varphi _{0}(x)$. The functions $\Psi
_{n}(x)$ are then readily deduced by using the aforementioned replacement
rule
\begin{equation}
\Psi _{n}(x)=\frac{N_{\Psi }}{\sqrt{n!2^{n}}}\left( \frac{\overline{\beta }+%
\overline{\delta }}{\overline{\alpha }+\overline{\gamma }}\right) ^{n/2}H_{n}%
\left[ \frac{x}{\sqrt{(\overline{\alpha }+\overline{\gamma })(\overline{%
\beta }+\overline{\delta })}}\right] \,e^{-\frac{1}{2}\,x^{2}\,\frac{%
\overline{\gamma }-\overline{\alpha }}{\overline{\gamma }+\overline{\alpha }}%
}.  \label{27}
\end{equation}%
We have now constructed our sets $\mathcal{F}_{\varphi }$ and $\mathcal{F}%
_{\Psi }$.

A special case of this general treatment was previously discussed in \cite%
{bag3}, where the pseudo-bosonic operators $a$ and $b$ were denoted as $%
A_{\theta }=\cos \theta \,c+i\sin \theta \,c^{\dagger }$ and $B_{\theta
}=\cos \theta \,c^{\dagger }+i\sin \theta \,c$, depending on a real parameter
$\theta \in I:=(-\frac{\pi }{4},\frac{\pi }{4})\setminus \{0\}$. Thus the
parameters in $T$ are identified as $\beta =\gamma =\cos \theta $ and $%
\delta =\alpha =-i\sin \theta $, clearly satisfying $\det (T)=1$ and the two
constraints in (\ref{25}) equal each other reducing to $\Re\left(
e^{2i\theta }\right) =\cos (2\theta )>0$ for $\theta \in I$. Furthermore,
the general solutions (\ref{25bis}) and (\ref{27}) for $\varphi _{n}(x)$ and $%
\Psi _{n}(x)$ simplify to
\begin{equation}
\varphi _{n}^{\theta }(x)=\frac{N_{\varphi }}{\sqrt{2^{n}\,n!}}%
\,\,H_{n}\left( e^{i\theta }x\right) \exp \left[ -\frac{1}{2}\,e^{2i\theta
}\,x^{2}\right] \qquad \Psi _{n}^{\theta }(x)=\frac{N_{\Psi }}{N_{\varphi }}%
\,\varphi _{n}^{-\theta }(x).
\end{equation}%
constructed in \cite{bag3} for the specific choice of parameters as given
above.

A direct computation shows that the two sets of functions $\mathcal{F}%
_{\varphi }$ and $\mathcal{F}_{\Psi }$ are indeed biorthogonal sets satisfying (%
\ref{FB35}). In previous analysis on ${\mathcal{D}}$-PBs \cite{baginbagbook} a
particular relevant role was played by the norms of $\varphi _{n}$ and $\Psi
_{n}$. For several concrete models it has been proved that $%
\lim_{n\rightarrow \infty }\Vert \varphi _{n}\Vert =\lim_{n\rightarrow
\infty }\Vert \Psi _{n}\Vert =\infty $, which is enough to conclude that $%
\mathcal{F}_{\varphi }$ and $\mathcal{F}_{\Psi }$ do not constitute bases
for $\mathcal{H}$, \cite{davies}, Lemma 3.3.3. However, this property does not exclude the possibility
that they are $\mathcal{D}$-quasi bases for some dense subspace $\mathcal{D}$
in $\mathcal{H}$. Thus we will next compute those limits in order to be able
to decide whether we encounter ${\mathcal{D}}$-PB  or no
PB at all and for which choices of the parameters in $T$ any of these
situations might occur.

We proceed by imposing some constraints on $T$ rather than considering the
complete generic case and compute $\Vert \varphi _{n}\Vert $ together with
the appropriate limit. This will make computations transparent at first.
Subsequently we investigate the consequences of relaxing some of the
constraints. In this manner we obtain unexpected and interesting conclusions
about the sets $\mathcal{F}_{\varphi }$ and $\mathcal{F}_{\Psi }$.

\subsection{Real valued GBT with constraint $\protect\alpha \protect\beta =%
\protect\gamma \protect\delta $}

\label{sect23}

We assume here that $T$ is real valued and its parameters are ordered as $%
\beta >\delta >0$ and that $\gamma >\alpha >0$. This choice guarantees that
the constraints in (\ref{25}) are automatically satisfied, whereas $\det
(T)=1$ must still be imposed. An explicit example for this choice of the
parameters is case d) provided after (\ref{25}). To compute $\Vert \varphi
_{n}\Vert $ we use the general formula
\begin{equation}
\int_{0}^{\infty }e^{-px^{2}}\,H_{n}(bx)\,H_{n}(cx)\,dx=\frac{2^{n-1}\,n!\,%
\sqrt{\pi }}{p^{(n+1)/2}}\,(b^{2}+c^{2}-p)^{n/2}\,P_{n}\left( \frac{bc}{%
\sqrt{p(b^{2}+c^{2}-p)}}\right) ,  \label{28}
\end{equation}%
which holds for all $p$ with positive real part \cite{prud} and $P_{n}(x)$
denotes the n-th Legendre polynomial. From (\ref{28}) and (\ref{25bis})
follows
\begin{equation}
\Vert \varphi _{n}\Vert ^{2}=\frac{\sqrt{\pi }|N_{\varphi }|^{2}}{2}\left(
\frac{\alpha +\gamma }{\beta +\delta }\right) ^{n}\left( \frac{\beta +\delta
}{\beta -\delta }\right) ^{\frac{n+1}{2}}\left( \frac{\gamma -\alpha }{%
\gamma +\alpha }\right) ^{\frac{n}{2}}P_{n}\left( \frac{1}{\sqrt{(\beta
^{2}-\delta ^{2})(\gamma ^{2}-\alpha ^{2})}}\right) .  \label{29}
\end{equation}%
We observe that $(\beta ^{2}-\delta ^{2})(\gamma ^{2}-\alpha ^{2})=1-(\alpha
\beta -\gamma \delta )^{2}$, which implies that the argument of the Legendre
polynomial is always greater or equal to one. It is suggestive to take, {to begin with, } $%
\alpha \beta =\gamma \delta $ as for that choice the expression in (\ref{29}%
) simplifies drastically due to the fact that $P_{n}(1)=1$ for all $n$. Thus
(\ref{29}) collapses to
\begin{equation}
\Vert \varphi _{n}\Vert ^{2}=\frac{\sqrt{\pi }|N_{\varphi }|^{2}}{2}\sqrt{%
\frac{\beta +\delta }{\beta -\delta }}\left( \frac{\gamma }{\beta }\right)
^{n},  \label{210}
\end{equation}%
and by using the aforementioned replacement rule we also obtain
\begin{equation}
\Vert \Psi _{n}\Vert ^{2}=\frac{\sqrt{\pi }|N_{\Psi }|^{2}}{2}\sqrt{\frac{%
\gamma +\alpha }{\gamma -\alpha }}\left( \frac{\beta }{\gamma }\right) ^{n}.
\label{211}
\end{equation}%
Then the conclusions are clear. We distinguish three cases:

\begin{equation}
\begin{array}{lll}
\gamma =\beta :\quad & \lim_{n\rightarrow \infty }\Vert \varphi _{n}\Vert =%
\text{const,\quad } & \lim_{n\rightarrow \infty }\Vert \Psi _{n}\Vert =\text{%
const,} \\
\gamma >\beta : & \lim_{n\rightarrow \infty }\Vert \varphi _{n}\Vert =\infty
, & \lim_{n\rightarrow \infty }\Vert \Psi _{n}\Vert =0, \\
\gamma <\beta : & \lim_{n\rightarrow \infty }\Vert \varphi _{n}\Vert =0, &
\lim_{n\rightarrow \infty }\Vert \Psi _{n}\Vert =\infty .%
\end{array}%
\end{equation}

For $\gamma =\beta $ we simply have $\varphi _{n}(x)=\Psi _{n}(x)N_{\varphi
}/N_{\Psi }$, since this choice also implies $\alpha =\delta $ and therefore
the GBT reduces to the standard Bogoliubov transformation with $a=b^{\dagger
}$. The two sets $\mathcal{F}_{\varphi }$ and $\mathcal{F}_{\Psi }$ essentially simply
collapse.

The situation $\gamma \neq \beta $ is more interesting. The fact that the
norms of the elements in $\mathcal{F}_{\varphi }$ and $\mathcal{F}_{\Psi }$
behave differently in the large $n$ asymptotic limit constitutes a new
result when compared with the many examples previously considered in the
literature. For instance, for the special case of the Swanson model, dealt
with in \cite{bag3}, it was found that both norms diverge $%
\lim_{n\rightarrow \infty }\Vert \varphi _{n}\Vert =\lim_{n\rightarrow
\infty }\Vert \Psi _{n}\Vert =\infty $.

The first clear conclusion is: \emph{Neither }$\mathcal{F}_{\varphi }$ \emph{%
nor} $\mathcal{F}_{\Psi }$ \emph{can be Riesz bases when} $\gamma \neq \beta
$. The reason is simple. For definiteness we take $\gamma >\beta $. Then,
since $\lim_{n\rightarrow \infty }\Vert \Psi _{n}\Vert =0$ one may imagine
the existence of an bounded operator $V$ and an orthonormal basis $\{e_{n}\}$
of $\mathcal{H}$ such that $\Psi _{n}=Ve_{n}$. However, due to the
uniqueness of the biorthogonal basis, we must necessarily have $\varphi
_{n}=(V^{-1})^{\dagger }e_{n}$. Now, since $\lim_{n\rightarrow \infty }\Vert
\varphi _{n}\Vert =\infty $, the operator $V^{-1}$ cannot be bounded, which
in turn implies our statement\footnote{%
We recall that in the definition of a Riesz basis the operator $V$ is
required to be invertible.}.

Our second conclusion is: $\mathcal{F}_{\varphi }$ \emph{and} $\mathcal{F}%
_{\Psi }$ \emph{constitute }\emph{two biorthogonal bases for }$ \Hil.$\emph{\ }%
Notice that this, of course, is not in contrast with the fact that they are not Riesz bases. The reason is because $\Vert \varphi _{n}\Vert $ and $\Vert
\Psi _{n}\Vert $ have a different asymptotic behavior, so that $\Vert
\varphi _{n}\Vert \Vert \Psi _{n}\Vert $ is uniformly bounded in $n$, { as required in} \cite%
{davies}. In fact, our second conclusion is not difficult to prove. Using $%
\det (T)=1$ and our constraint $\alpha \beta =\gamma \delta $ we can
eliminate $\alpha $ and $\gamma $ from the expressions $\varphi _{n}(x)$ in (%
\ref{25bis}) and $\Psi _{n}(x)$ in (\ref{27}) and express them entirely in
terms of the parameters $\beta $ and $\delta $
\begin{equation}
\varphi _{n}(x)=\frac{N_{\varphi }}{\sqrt{2^{n}\,n!}}\left( \frac{1}{\beta
^{2}-\delta ^{2}}\right) ^{n/2}H_{n}\left( \sqrt{\frac{\beta -\delta }{\beta
+\delta }}\,x\right) e^{-\frac{1}{2}x^{2}\frac{\beta -\delta }{\beta +\delta
}}
\end{equation}%
and
\begin{equation}
\Psi _{n}(x)=\frac{N_{\Psi }}{\sqrt{2^{n}\,n!}}\left( \beta ^{2}-\delta
^{2}\right) ^{n/2}H_{n}\left( \sqrt{\frac{\beta -\delta }{\beta +\delta }}%
\,x\right) e^{-\frac{1}{2}x^{2}\frac{\beta -\delta }{\beta +\delta }},
\end{equation}%
with $\overline{N}_{\varphi }N_{\Psi }=\sqrt{(\beta -\delta )/(\beta +\delta
)\pi }$. Note that by our initial assumption the arguments of all the square
roots are positive. Then we have $\Psi _{n}(x)=\left( \beta ^{2}-\delta
^{2}\right) ^{n}\varphi _{n}(x)N_{\Psi }/N_{\varphi }$ for all $n\geq 0$,
i.e. the two sets only differ by a constant, albeit $n$-dependent, factor.
Taking now a generic function $f(x)\in \mathcal{L}^{2}(\mathbb{R)}$ simple
manipulations show that
\begin{equation}
\sum_{n=0}^{\infty }\left\langle \varphi _{n},f\right\rangle \Psi
_{n}(x)=\sum_{n=0}^{\infty }\frac{\overline{N}_{\varphi }N_{\Psi }\sqrt{\pi }%
}{\mu }\left( \int_{\mathbb{R}}e_{n}(s)f_{\mu }(s)ds\right) e_{n}(t),
\end{equation}%
where we have introduced the positive quantity $\mu =\sqrt{(\beta -\delta
)/(\beta +\delta )}$, the variable $t=x\mu $, the shorthand notation $f_{\mu
}(s)=f\left( s/\mu \right) $ and the function $e_{n}(s)=H_{n}(s)e^{-\frac{1}{%
2}s^{2}}/\sqrt{2^{n}\,n!\sqrt{\pi }}$, which all together (i.e. for $%
n=0,1,2,\ldots $) form an orthonormal basis for $\mathcal{L}^{2}(\mathbb{R)}$%
. This implies that
\begin{equation}
\sum_{n=0}^{\infty }\left\langle \varphi _{n},f\right\rangle \Psi
_{n}(x)=f_{\mu }(t)=f(x),
\end{equation}%
which is what we had to prove. Therefore $\mathcal{F}_{\Psi }$ is a basis.
Analogously, we can show that $\mathcal{F}_{\varphi }$ is a basis too.
Moreover, they are clearly both $\mathcal{H}$-quasi bases.

\noindent \textbf{Remark:} This is not very different from what happens if
we start with a generic orthonormal basis $\mathcal{E}=\{e_{n}\}$ of $%
\mathcal{H}$ and construct out of it two sets $\mathcal{F}_{\varphi }=\{%
\varphi _{n}=\lambda _{n}e_{n}\}$ and $\mathcal{F}_{\Psi }=\{\Psi _{n}=\lambda
_{n}^{-1}e_{n}\}$, using a sequence $\{\lambda _{n}\}$ of non zero complex
numbers. Then, if for instance $|\lambda _{n}|\leq M$, for some $0<M<\infty $%
, with divergent $\lambda _{n}^{-1}$, it is clear that: (i) $\mathcal{F}%
_{\varphi }$ and $\mathcal{F}_{\Psi }$ are not Riesz bases, (ii) They are
biorthogonal and they are both bases for $\mathcal{H}$ and (iii) they are
both $\mathcal{H}$-quasi bases.

{It is easy to understand why, when $\det(T)=1$ and $\alpha \beta =
\gamma \delta $, the sets $\F_\varphi$ and $\F_\Psi$ essentially collapse and became bases for $\Hil$. The reason is that, under these assumptions on the coefficients, the number operator $N=ba$ is self-adjoint\footnote{It might be useful to observe that condition $\gamma=\beta$ needs not to be satisfied, here, since it is a sufficient but not a necessary condition to have $N=N^\dagger$.}. In fact, we find that
$$
N=\frac{1}{\beta^2-\delta^2}\,\left(\beta c-\delta c^\dagger\right)^\dagger \left(\beta c-\delta c^\dagger\right).
$$
So, we are just working with a sort of rescaled self-adjoint harmonic oscillator. This will not be so in the next Section, where something completely different will be deduced.

}

\subsection{Removing constraint $\protect\alpha \protect\beta =\protect%
\gamma \protect\delta $}

\label{sect24}

From the previous subsection it appears at first sight that the constraint $%
\alpha \beta =\gamma \delta $ only facilitated our computations. We will now
demonstrate that it actually describes a very special situation and when it
is relaxed the properties of our PBs change severely, i.e. we find that
\emph{neither} $\mathcal{F}_{\varphi }$ \emph{nor} $\mathcal{F}_{\Psi }$
\emph{constitute bases for} $\mathcal{H}$ \emph{when the coefficients in the
GBT are such that} $\alpha \beta \neq \gamma \delta $. The proof of this
claim makes use of the fact that in this case the argument of the Legendre
polynomial in (\ref{29}) is always larger or equal to one. Then, to deduce
the asymptotic behavior of $\Vert \varphi _{n}\Vert $, {where $\varphi_n(x)$ are now those in (\ref{25bis})}, for large $n$ we can
employ the following formula, see \cite{szego},
\begin{equation}
P_{n}(x)\simeq \frac{1}{\sqrt{2\pi n}}\,\frac{1}{(x^{2}-1)^{1/4}}\left\{ x+%
\sqrt{x^{2}-1}\right\} ^{n+1/2},  \label{212}
\end{equation}%
which holds if $x>1$. The asymptotic behavior of (\ref{29}) and the
analogous one for $\Psi _{n}$
\begin{equation}
\Vert \Psi _{n}\Vert ^{2}=\frac{\sqrt{\pi }|N_{\Psi }|^{2}}{2}\left( \frac{%
\delta +\beta }{\gamma +\alpha }\right) ^{n}\left( \frac{\gamma +\alpha }{%
\gamma -\alpha }\right) ^{\frac{n+1}{2}}\left( \frac{\beta -\delta }{\beta
+\delta }\right) ^{\frac{n}{2}}P_{n}\left( \frac{1}{\sqrt{(\beta ^{2}-\delta
^{2})(\gamma ^{2}-\alpha ^{2})}}\right) ,  \label{213}
\end{equation}%
are described by
\begin{equation}
\Vert \varphi _{n}\Vert ^{2}\simeq A^{\varphi }\frac{x^{n}}{\sqrt{n}},\qquad
\Vert \Psi _{n}\Vert ^{2}\simeq A^{\Psi }\frac{y^{n}}{\sqrt{n}}.  \label{214}
\end{equation}%
We introduced here the quantities
\begin{equation}
x:=\frac{1+|\alpha \beta -\gamma \delta |}{\beta ^{2}-\delta ^{2}},\qquad y:=%
\frac{1+|\alpha \beta -\gamma \delta |}{\gamma ^{2}-\alpha ^{2}},
\end{equation}%
and%
\begin{equation}
A^{\varphi }:=\frac{A\sqrt{\pi }|N_{\varphi }|^{2}}{2}\sqrt{\frac{\beta
+\delta }{\beta -\delta }},\qquad A^{\Psi }=\frac{A\sqrt{\pi }|N_{\Psi }|^{2}%
}{2}\sqrt{\frac{\gamma +\alpha }{\gamma -\alpha }},
\end{equation}%
with{
\begin{equation}
A=\frac{1}{\sqrt{2\pi }}\frac{1}{(s^{2}-1)^{1/4}}\left[ s+\sqrt{s^{2}-1}%
\right] ^{1/2},
\end{equation}%
where $s=\frac{1}{\sqrt{(\beta ^{2}-\delta
^{2})(\gamma ^{2}-\alpha ^{2})}}$.
}

From (\ref{214}) it is clear that $\lim_{n\rightarrow \infty }\Vert \varphi
_{n}\Vert =0$ for $0<x\leq 1$ and that it diverges when $x>1$. Analogously,
we have $\lim_{n\rightarrow \infty }\Vert \Psi _{n}\Vert =0$ for $0<y\leq 1$
and $\lim_{n\rightarrow \infty }\Vert \Psi _{n}\Vert =\infty $ when $y>1$.
Therefore, at a first sight, we might expect to recover the same situation
as in the previous section where the product $\Vert \varphi _{n}\Vert \Vert
\Psi _{n}\Vert $ turned out to be independent of $n$, see (\ref{210}) and (%
\ref{211}). In particular, this product was uniformly bounded in $n$, which
is a necessary (but not sufficient) condition for $\mathcal{F}_{\varphi }$
and $\mathcal{F}_{\Psi }$ to be bases for $\mathcal{H}$, see \cite{davies}, Lemma 3.3.3.
In contrast, we will see that this condition is never satisfied in the
present setting. The proof of this behavior is indeed simple. From (\ref%
{214}) follows that
\begin{equation}
\Vert \varphi _{n}\Vert ^{2}\Vert \Psi _{n}\Vert ^{2}=A^{\varphi }A^{\Psi }%
\frac{(xy)^{n}}{n}=A^{\varphi }A^{\Psi }\frac{1}{n}\left( \frac{1+|\alpha
\beta -\gamma \delta |}{1-|\alpha \beta -\gamma \delta |}\right) ^{n},
\end{equation}%
which diverges with $n\rightarrow \infty $ whenever $\alpha \beta -\gamma
\delta \neq 0$. Thus in this case we recover the behavior already
encountered for the special case of the Swanson model, where both $\mathcal{F%
}_{\varphi }$ and $\mathcal{F}_{\Psi }$ were shown not to be bases.
Nonetheless, we are left with the possibility that they are ${\mathcal{D}}$%
-quasi bases. And indeed, this is what happens.

To prove the latter, we begin by introducing the set
\begin{equation}
{\mathcal{D}}=\left\{ f(x)\in \mathcal{L}^{2}(\mathbb{R)}:\,e^{\frac{1}{2}%
x^{2}|\alpha \beta -\gamma \delta |}f(x)\mathbb{\in }\mathcal{L}^{2}(\mathbb{%
R)}\right\} .  \label{add1}
\end{equation}%
This set is dense in $\mathcal{L}^{2}(\mathbb{R)}$, since it contains the
set $D(\mathbb{R)}$ of the $C^{\infty }$ functions with bounded support.
Moreover, if $\alpha \beta \neq \gamma \delta $, it clearly does not
coincide with $\mathcal{L}^{2}(\mathbb{R)}$. Now, if $f(x)$ and $g(x)$
belong to ${\mathcal{D}}$, we can check that
\begin{equation}
\left\langle f,\varphi _{n}\right\rangle =A_{n}\sqrt{(\alpha +\gamma )(\beta
+\delta )}\int_{\mathbb{R}}\overline{f_{1}(x)}\,H_{n}(x)e^{-x^{2}/2}dx
\end{equation}%
and
\begin{equation}
\left\langle \Psi _{n},g\right\rangle =B_{n}\sqrt{(\alpha +\gamma )(\beta
+\delta )}\int_{\mathbb{R}}H_{n}(x)e^{-x^{2}/2}{g_{1}(x)}\,dx,
\end{equation}%
where, to simplify the notation, we have introduced
\begin{equation}
A_{n}=\frac{N_{\varphi }}{\sqrt{2^{n}\,n!}}\left( \frac{\alpha +\gamma }{%
\beta +\delta }\right) ^{n},\qquad B_{n}=\frac{\overline{N_{\Psi }}}{\sqrt{%
2^{n}\,n!}}\left( \frac{\beta +\delta }{\alpha +\gamma }\right) ^{n},
\end{equation}%
and
\begin{equation*}
f_{1}(x)=f\left( x\sqrt{(\alpha +\gamma )(\beta +\delta )}\right) \,e^{\frac{%
1}{2}x^{2}(\gamma \delta -\alpha \beta )},~~g_{1}(x)=g\left( x\sqrt{(\alpha
+\gamma )(\beta +\delta )}\right) \,e^{\frac{1}{2}x^{2}(\alpha \beta -\gamma
\delta )}.
\end{equation*}%
Notice that, since both $f(x)$ and $g(x)$ are taken into ${\mathcal{D}}$, $%
f_{1}(x)$ and $g_{2}(x)$ are square integrable, even if $\gamma \delta \neq
\alpha \beta $. Now, using again the orthonormal eigenfunctions of the
harmonic oscillator
\begin{equation}
e_{n}(x)=\frac{1}{\sqrt{2^{n}\,n!\sqrt{\pi }}}\,H_{n}(x)\,e^{-x^{2}/2},
\end{equation}%
for $n=0,1,2,\ldots $, we deduce that
\begin{eqnarray}
\sum_{n=0}^{\infty }\left\langle f,\varphi _{n}\right\rangle \left\langle
\Psi _{n},g\right\rangle  &=&\sqrt{\pi }\,(\alpha +\gamma )(\beta +\delta
)N_{\varphi }\overline{N_{\Psi }}\sum_{n=0}^{\infty }\left\langle
f_{1},e_{n}\right\rangle \left\langle e_{n},g_{1}\right\rangle  \\
&=&\sqrt{\pi }\,(\alpha +\gamma )(\beta +\delta )N_{\varphi }\overline{%
N_{\Psi }}\left\langle f_{1},g_{1}\right\rangle =\left\langle
f,g\right\rangle .
\end{eqnarray}%
Analogously one can prove that $\sum_{n=0}^{\infty }\left\langle f,\Psi
_{n}\right\rangle \left\langle \varphi _{n},g\right\rangle =\left\langle
f,g\right\rangle $, for all $f,g\in {\mathcal{D}}$. The conclusion is that $%
\mathcal{F}_{\varphi }$ and $\mathcal{F}_{\Psi }$, \underline{even if they are not
bases for $\mathcal{H}$},  are ${\mathcal{D}}$-quasi bases. This is what
happens also for the aforementioned special case of the Swanson model,
which, however, differs from the situation considered here since in that
case some of the coefficients were complex valued. Therefore, apparently,
having real or complex parameters in $T$ does not prevent the sets $\mathcal{%
F}_{\varphi }$ and $\mathcal{F}_{\Psi }$ to be ${\mathcal{D}}$-quasi bases,
while among all the real possibilities, there exists just a particular family of choices
which reproduces biorthogonal bases for $\mathcal{H}$. Other choices of real parameters produce not bases, but $\D$-quasi bases.

\section{Conclusions}

In this manuscript we have studied the relations between GBTs and ${\mathcal{%
D}}$-PBs. We have found the interesting possibility that GBTs may produce
examples of biorthogonal sets that are in addition bases for a Hilbert space
$\mathcal{H}$. When the map $T$ in (\ref{21}) that defines the GBT is taken
to be real valued and its parameters are ordered as $\beta >\delta >0$ and $%
\gamma >\alpha >0$ we found two qualitatively different situation. Imposing
in addition the constraint $\alpha \beta =\gamma \delta $ we found the
hitherto unobserved behavior that the norms of the vectors in sets of the eigenvectors of two related number operators, $\mathcal{F}_{\varphi }$ and $%
\mathcal{F}_{\Psi }$, respectively, possess different types of asymptotic
behavior. We concluded from this that they do not form  Riesz bases, but
still they constitute two biorthogonal bases for $\Hil$, and $\Hil$-quasi bases as a consequence. In contrast, when we relax the
constraint and take $\alpha \beta \neq \gamma \delta $ instead we proved
that neither $\mathcal{F}_{\varphi }$ nor $\mathcal{F}_{\Psi }$ are bases
for $\mathcal{H}$. Nonetheless, even in this case the sets $\mathcal{F}%
_{\varphi }$ and $\mathcal{F}_{\Psi }$ are still ${\mathcal{D}}$-quasi bases
for a suitable ${\mathcal{D}}$ dense in $\mathcal{H}$ as specified in (\ref%
{add1}). The latter behavior was previously observed for specific complex
choices of the parameters in $T$ related to a particular version of the
Swanson model. In fact, see \cite{baginbagbook}, in this case the sets $\mathcal{F%
}_{\varphi }$ and $\mathcal{F}_{\Psi }$ are $\mathcal{G}$-quasi bases, where
$\mathcal{G}$ is  the linear span of the $e_{n}(x)$'s, which is
obviously dense in $\mathcal{H}$.

One may also consider the reverse construction, i.e. the possibility of
constructing a GBT out of a family of ${\mathcal{D}}$-PBs. Indeed, this is
either trivial as in the example provided with operators $A_{\theta }$ and $%
B_{\theta }$, where one simply has to read off the values for the
complex-valued parameters $\alpha $, $\beta $, $\gamma $ and $\delta $ to
define the map $T$ that represents the GBT or it is not possible at all. The
latter case emerges for instance when the ${\mathcal{D}}$-PBs are
constructed by adding complex constants to the standard representations of
the operators $c$ and $c^{\dagger }$, see for instance \cite{bagfring}.
Clearly such a construction can not be cast into the form of a GBT of the
form as specified in (\ref{21}). In summary, not all GBT correspond to PBs
and vice versa not all versions PBs may be cast into the form of a GBT.

There are clearly some challenges left. Obviously to complete the picture it
would be interesting to study the behavior for the remaining choices of $%
\alpha $, $\beta $, $\gamma $ and $\delta $ not covered in our treatment. In
addition, it would be interesting to study these aspects in more complicated
models based on GBT, such as the non-Hermitian Hamiltonians of Lie algebraic
type investigated in \cite{andreas1}.

\section*{Acknowledgements}

F.B. acknowledges support by the University of Palermo and GNFM.

\end{document}